\documentclass[12pt]{article}
\usepackage{epsf}
\usepackage{cite}
\def\beq{\begin{eqnarray}}
\def\eeq{\end{eqnarray}}

\def\Mpl{M_{\rm Pl}}
\def\Rbr{R}
\def\Rbu{{\cal R}}
\def\M{M_*}
\def\gn{G_N}

\begin{document}
\begin{flushright}
NYU-TH/02/12/1 \\
\end{flushright}

\vspace{0.1in}
\begin{center}
\bigskip\bigskip
{\large \bf  The Accelerated Universe and the Moon}

\vspace{0.4in}      

{Gia Dvali, Andrei Gruzinov, and Matias Zaldarriaga}
\vspace{0.1in}

{\baselineskip=14pt \it 
Center for Cosmology and Particle Physics\\[1mm]
}{\baselineskip=14pt \it 
Department of Physics, New York University,  
New York, NY 10003\\[1mm]
}
\vspace{0.2in}
\end{center}

\vspace{0.9cm}
\centerline{\bf Abstract}
 
Cosmologically motivated theories that explain small acceleration rate of the Universe
via modification of gravity at very large, horizon or super-horizon distances, can be tested by precision gravitational measurements at much shorter
scales, such as the Earth-Moon distance. Contrary to the naive expectation the predicted  corrections to the Einsteinian metric near  gravitating sources are so significant that might fall within sensitivity of the proposed Lunar Ranging experiments. 
The key reason for such corrections is the van Dam-Veltman-Zakharov discontinuity
present in linearized versions of all such theories,  and its subsequent  absence at the non-linear level ala Vainshtein. 


\newpage
\section{Generalities}

Recent observations suggest that the Universe is accelerating on the scales of the present
cosmological horizon\cite{acceleration}.  This indicates that, either there is a small vacuum energy (or some  ``effective" vacuum energy), or that conventional laws of Einstein gravity get modified at very large distances, imitating a small cosmological constant
\cite{dgp,dgs,addg}.
 The first possibility is unnatural in the view of quantum field theory\footnote{An alternative explanation could  be provided by Anthropic approach\cite{anthropic}}, since the required value of vacuum energy $\sim 10^{-12}$GeV$^4$ is unstable under quantum corrections. This
unnaturalness goes under the name of the Cosmological Constant Problem. 
  
In this respect the second approach, of modifying gravity in far infrared can be more promising since it is perturbatively stable under quantum corrections. The unnaturally small value of the
vacuum energy is replaced by the idea that laws of conventional gravity break down at very large distances, beyond a certain crossover scale  $r_c$.  The value of $r_c$ is 
{\it perturbatively-stable} ie. it  does not suffer from cut-off sensitive corrections experienced by vacuum energy.   
This modification at the level of linearized effective four-dimensional equation for the freely propagating metric fluctuations $h_{\mu\nu}(x)$ about the flat background can be described in  the following general way\cite{dgp,dgs,addg,es}:
\begin{equation}
\label{modify}
 (\nabla^2 \, + {1 \over r_c^2} f(r_c^2\nabla^2)\, ) \, h_{\mu\nu} \, = \,  0
\end{equation}
Here $\nabla^2 \, = \, \nabla_{\mu}\nabla^{\mu}$ denotes four-dimensional 
d'Alambertian, and  $f(r_c^2\nabla^2)/r_c^2$ is an operator that dominates over $\nabla^2$
only for momenta $q \ll \, 1/r_c$.
One way to think  of this modification is that the effective gravitational coupling (effective Newton's ``constant")
\begin{equation}
\label{filter}
G_{eff} \,  = \, { 1 \over 8\pi \Mpl^2}   \left [1 \, + {f(r_c^2\nabla^2)\, \over r_c^2\nabla^2} \right ]^{-1}
\end{equation}
becomes dependent on the wave-length.
We will assume that the function $f(r_c^2\nabla^2)$ is
expandable in power series of $\nabla^2$.

Depending on the precise structure, such models can be divided in the following three categories.  
The first category is the model\cite{dgp}, in which Newtonian gravity turns into the five dimensional
$1/r^2$-potential at distances $r \gg r_c$. This effect is due to existence of an infinite-volume
flat extra dimension to which gravity ``leaks" from the 3-brane, where the conventional particles
live. So a brane-observer sees effectively four-dimensional theory of gravity, in which the effective 4D graviton is unstable, but with an arbitrarily large  lifetime $\tau \sim r_c$,  over which it decays into five-dimensional continuum of states.  

 Although extra dimensions play crucial role in formulating the manifestly generally-covariant
theory, for our purposes we can equally well use purely 4D language.   For the brane observer
interested in metric on the brane created by brane sources, there is an effective four-dimensional
description in terms of a single four-dimensional graviton $h_{\mu\nu}$.  Not surprisingly, this
effective 4D theory is non-local, but non-local terms only dominate in far infrared. Moreover, if one is interested in estimates of sub-leading corrections  to Einstein metric near the gravitating sources, the 4D language is simpler.  At the linearised level this 4D theory is given by (\ref{modify}) with
\cite{dgs}
\begin{equation}
\label{fdgp}
f(r_c^2\nabla^2) \, =  \, r_c \sqrt{-\nabla^2}
\end{equation}
In this theory the smallness of cosmological constant  is not explained. However, if one postulates that vacuum energy is zero due to some other reason, the model explains  the accelerated expansion of the Universe without any need of dark energy\cite{cedric,ddg}.  The reason, as discovered in \cite{cedric}, is that  in \cite{dgp} the effective Friedmann equation gets modified by additional  powers of Hubble parameter $H$, that dominate for very low curvatures  (late times). In the leading order the resulting equation can be parameterized as\cite{cedric,ddg}
\begin{equation}
\label{hubble}
H^2 \, - \, {H\over r_c} \,  = \,  {\rho \over 3 \Mpl^2}.
\end{equation}
At late times, this equation has a self-accelerating cosmological branch with $H \, = \, {1 \over r_c}$.
Observation fix the crossover scale 
to be $r_c \sim 10^{28}$cm. Confronted with minimal models where dark energy is a pure vacuum
energy, the above model also contains just one parameter  ($r_c$ versus the vacuum energy), but the difference is that the value of $r_c$ is insensitive to quantum loops.  

 In  two other classes of models\cite{dgs, addg},  one can address the  cosmological constant problem .  Models of \cite{dgs} are based on higher-dimensional generalization of \cite{dgp} with
$N\, > \, 1$ extra dimensions \cite{dg}.  As a result, the feature of 4D Newtonian gravity turning into
the high-dimensional $1/r^{1 + N}$-potential for $r \gg r_c$  is  shared by  $N > 1$ theories. 
However,  having more than one extra dimensions proved to be crucial for
cosmological constant. In these theories the large 4D vacuum energy (${\cal E}$)  curves the four-dimensional space only mildly resulting in an accelerated expansion, with the rate  inversely proportional to a power of  ${\cal E}$. In four-dimensional language,  gravity is modified in such a way that sources of wave-length $\gg r_c$ gravitate very weakly. Thus although vacuum energy is huge, it actually does not curve the space and can lead to an observable small acceleration.

 Finally there is a third class \cite{addg, es}, which does not refer to any underlying high-dimensional theory, but is rather based on infrared modification of gravity directly in four-dimensions.   In this models,  $G_{eff}$ is postulated to act as a high-pass filter, so that as in
\cite{dgs}  sources of wavelength $\gg r_c$ gravitate extremely  weakly. 
 
In the above models, value of $r_c$ may of may not be fixed. For instance, in the models of \cite{dgs} it is actually fixed to be $r_c \sim 10^{28}$ cm , due to the current lower limits on Newtonian $1/r$ gravity.

 The common feature of all the above theories, is that  beyond the
 crossover scale $r_c$  Newtonian $1/r$ gravity gets substantially modified. 
But,  at short distances  (near the gravitating sources), all the prediction of
 Einstein gravity are reproduced up to some small corrections.  These corrections are
the central interest of the present work.

 Given the fact that crossover scale is
$r_c$, one could naively conclude that deviations are suppressed by powers of ${r\over r_c}$.
If this were the case, then for $r_c \sim 10^{28}$cm, the corrections would be completely
negligible at any distance
of potential precision measurements.

  Interestingly, the story is very different. Recall that near any gravitating object there is a second (much smaller scale) in the problem, the gravitational radius of the body $r_g$. 
It was shown in 
\cite{gruzinov} that leading corrections to the gravitational potential $\Psi$ in the model 
of \cite{dgp} are strongly enhanced by
inverse powers $r_g$ and are given by
\begin{equation}
\label{andrei}
\epsilon \, = \, { \delta \Psi \over \Psi}  \sim  {r \over r_c}\sqrt{r\over r_g}    
\end{equation}
The aim of the present work is to show that in a large class of theories that modify gravity
beyond some horizon or super-horizon size distance $r_c$,  there
are corrections of the type \cite{gruzinov} that penetrate at much shorter scales $r_* \ll r_c$, and
could be potentially measurable. 
These corrections can be detected in precision gravitational measurements in systems that
are much smaller than $r_c$,  e.g., such as the Earth-Moon system. In most interesting cases, such as the expression  (\ref{andrei}),  they are on the border line of existing measurements, and thus can be detected if the existing sensitivities are improved.

 The key reason for such an unexpected behavior is that the graviton in the above-discussed theories has extra polarization that also couples to the conserved energy-momentum source, and
mediates a scalar-type force. As a result, at the {\it linearized} level gravity is of that scalar-tensor
type.  Therefore whenever linearized approximation is valid the predictions of the above theory
differ from those of Einstein gravity by a finite amount, no matter how small is the ratio $r/r_c$.
 
This effect was originally pointed out in the framework of linearized theory of massive gravity\cite{vdvz}, and goes under the name of van Dam - Veltman - Zakharov discontinuity
(vDVZ).  It lead vDVZ to the conclusion that massive theories of gravity are ruled out. 
Later the similar effect  was observed  in the linearized version of the generally-covariant
model of\cite{dgp}.

 The crucial point\cite{Arkady}, however, is that discontinuity is an artifact of the linearized approximation, and  is cured by non-linear corrections. In reality, the solution is continuous 
in the limit $1/ r_c \rightarrow 0$.  
This was originally suggested by Vainshtein  in\cite{Arkady}\footnote{
On AdS space vDVZD can be absent already at the linear level, if graviton mass is smaller
than the AdS curvature \cite{Higuchi, Kogan, Porrati}. The roots of this phenomenon can again be
attributed to the effect of  \cite{Arkady}. Such situation probably is not experimentally interesting interesting, and won't be discussed here.}  and later confirmed by explicit 
fully-nonlinear analysis in generally covariant model of \cite{dgp},
 both in cosmological solutions\cite{ddgv}, as well as for  localized sources, such as
cosmic strings\cite{lue} and Schwarzschild \cite{gruzinov, masimo}.  
 These studies uncover the same persistent pattern.  Near the sources the solutions are
arbitrarily close to those of Einstein gravity, and continuous for large $r_c$.  However,
the corrections set in at distances, (or time scales) much smaller than $r_c$, which makes them
potentially observable even if $r_c$ is very large (e.g. horizon size) (the relevance of this corrections for the orbit of Jupiter was pointed out in \cite{gruzinov}).
This makes us think that
analogous behavior must take place in any theory with  infrared modification of gravity of the form in (\ref{modify}).
The key message of our analysis is that in all such theories of interest vDVZ discontinuity indeed disappears at the non-linear level in accordance to \cite{Arkady,ddgv,lue,gruzinov},  and standard predictions of Einstein gravity are recovered near the gravitating sources, where non-linearities are important. 
But as in \cite{gruzinov} for any given localized source, there exists
a distance $r_*$ for which the non-linearities are unimportant and thus the ``wrong predictions" take over. 

We give a qualitative prescription to estimate the leading order corrections to the Einstein metric using the form of the function
$f(r_c^2\nabla^2)$ in theories interest. Using this prescription one can derive
observational constraints on such theories. 

 The rest of the paper is organized as follows. In section 2, we briefly summarize our theoretical results and in section 3  we discuss their relevance for observations, through the anomalous
perihelion precession of orbits of planets. The reader interested in purely the phenomelogical
applications of our results can dismiss the rest of the paper which is devoted to more theoretical considerations. In section 4 we discuss the role of extra graviton polarization in creating the observable effects at intermediate scales in theories with infrared-modified gravity. 
In section 5 we briefly discuss the applicability range of our results.

\section{Results}

 Here we shall briefly  summarize our results. We will be interested in a class of
generally-covariant theories, containing exclusively spin-2 states, for which the linearized perturbations about the flat space satisfy: 
\begin{equation}
\label{modf}
\left (\, \nabla^2 \, + \, {1 \over r_c^2} f(r_c^2\nabla^2) \,
\right ) \, h_{\mu\nu} \, = -\, {1 \over M_{Pl}^2} \, \left \{ T_{\mu\nu} -{\beta \over 2}\eta_{\mu\nu} 
T^\alpha_\alpha   \right \}\, + \, \cdots
\end{equation}
where  the ellipsis stands for some derivative-dependent terms that vanish in convolution with conserved sources.  The over-all coefficient on r.h.s. was absorbed in redefinition of $M_{Pl}$. 
In all theories of interest  $\beta \,  \neq \, 1$, as opposed to the case of Einstein theory of massless  graviton with two polarizations, 
in which $\beta \, = \, 1$. Thus the  theory exhibits the analog of vDVZ discontinuity. 

Then our analysis suggests the following. 

As suggested by Vainshtein,  vDVZ is cured at the non-linear level.  That is, the solution are continuous in the limit $r_c \rightarrow \infty$ near the sources.
For any gravitating source, with gravitational radius $r_g \ll r_c$ there is an important intermediate scale in the problem
\begin{equation}
\label{scale}
r_g \, \ll \,  r_* \,  \ll \, r_c 
\end{equation}
where $r_* $ is determined from the following equation
\begin{equation}
\label{rstar}
\sqrt{r_g\over r_*} \, \sim  \, \left ( {r_*^2 \over r_c^2}\right ) \,  f\left ( {r_c^2 \over r_*^2}\right ) 
\end{equation}

The various distance scales work as follows. For $r \ll r_*$ the metric produced by the source is nearly Schwarzchild, with the leading correction to the metric that can be estimated as
\begin{equation}
\label{corrections}
\delta \Psi \sim {\sqrt{r_g\, r^3} \over r_c^2}  \,  f\left ( {r_c^2 \over r^2}\right ) 
\end{equation}
This expression, for the particular form of the $f$-function given by  (\ref{fdgp}),  correctly reproduces the result of \cite{gruzinov} given in (\ref{andrei}),  found by solving the full nonlinear high-dimensional equation of \cite{dgp}. 

As will be discussed below, the above corrections can play the crucial role for testing theories in question in precision gravitational measurements.

In the interval $r_* \ll r \ll r_c$, the linearized approximation is valid. So  gravity is still
$1/r$, but is of scalar-tensor type.  Note that the term ``scalar-tensor" only refers to the
similarity in the tensor structure, rather than to existence of an independent spin-0 state in the theory. Instead the ``scalar" admixture comes from the extra polarization of spin-2 graviton. 

Finally, for the distances $ r \gg r_c$, gravity becomes unconventional, with the potential given by
\begin{equation}
\label{farpot}
\Psi(r) \sim \int d^3q \, {r_c^2 \over f(r_c^2q^2)}\,  {\rm e}^{-iqr} 
\end{equation}
where $q$ is the three-momentum.

\section{Anomalous Perihelion Precession}

 The considered class of theories predict  slight modification of  the gravitational
potential of a massive body at observed distances according to (\ref{corrections}). This modifications can be observable in
the experiments that are sensitive to anomalous perihelion precession of planets. 
 Let $\epsilon$ be the
fractional change of the gravitational potential
\begin{equation}
\epsilon \equiv {\delta \Psi \over \Psi },
\end{equation}
where $\Psi=-GM/r$ is the Newtonian potential. The anomalous perihelion
precession (the perihelion advance per orbit due to gravity
modification) is
\begin{equation}
\delta \phi =\pi r (r^2(r^{-1}\epsilon )')',
\end{equation}
where $'\equiv d/dr$.

Let us apply this to the model of \cite{dgp}.
In this theory, \cite{gruzinov}
\begin{equation}
\epsilon \, = \, - \, \sqrt{2}r_c^{-1}r_g^{-1/2}r^{3/2}.
\end{equation}
The numerical coefficient deserves some clarification. The above coefficient was derived in
\cite{gruzinov} on Minkowski background. However, non-linearities created by cosmological expansion can further correct the coefficient. One would expect these corrections to scale as
powers of $r_cH$, where $H$ is the observed value of the Hubble parameter. On the accelerated branch \cite{cedric}, as it's obvious from  (\ref{hubble}),  $H \, \sim \, 1/r_c$ and thus,  
one would expect the corrections to be of order one. 
Recently, a very interesting fact was pointed out by Lue and Starkman\cite{ls}\footnote{We
thank these authors for sharing their preliminary results with us.}, that the cosmological background only affects the sign of the coefficient. The sign depends on  the particular cosmological branch. It is negative  for the standard cosmological branch, and positive for the self-accelerated one. We will restrict ourselves to order of magnitude estimate, but the sign will be very important if the effect is found, since according to\cite{ls} it could give information about the cosmological branch.

We get
\begin{equation}
\delta \phi =(3\pi /4)\epsilon.
\end{equation}
Numerically, the gravitational radius of the Earth is $r_g=0.886$cm, the
Earth-Moon distance is $r=3.84\times 10^{10}$cm, the gravity
modification parameter that gives the observed acceleration without dark
energy $r_c=6$ Gpc. We get the theoretical precession
\begin{equation}
\delta \phi = 1. 4 \times 10^{-12}.
\end{equation}
This is to be compared to the accuracy of the precession measurement by
the lunar laser ranging. Today the accuracy is $\sigma _\phi
=2.4\times 10^{-11}$ and no anomalous precession is detected at this
accuracy \cite{dickey}. In the future a tenfold improvement of the accuracy is expected \cite{adelberger}.

As noted in \cite{ls}, anomalous Martian precession is also of interest for testing \cite{dgp} with cosmologically intersting values of $r_c$. For $r_c=6$ Gpc we get $\delta \phi _{Mars}= 4 \times 10^{-11}$. This is to be compared to the accuracy of $\sigma _{\phi _{Mars}}= 9 \times 10^{-11}$ which might become possible as a result of the Pathfinder mission \cite{nordt}.

Let us now consider some generalizations.
In \cite{dgp}  $f(r_c^2\nabla^2)$ has the form (\ref{fdgp}). 
Consider now the minimal modification of the form
\begin{equation}
f(r_c^2\nabla^2) \, = \,  (r_c^2 \nabla^2)^{{1 - \gamma \over 2}}.
\label{gamma}
\end{equation}
According to (\ref{corrections}) this modification of the graviton propagator gives
the fractional change of the gravitational potential analogous to (3):
\begin{equation}
\epsilon \sim r_c^{-1-\gamma }r_g^{-1/2}r^{3/2+\gamma }.
\end{equation}
The corresponding anomalous lunar perihelion precession is
\begin{equation}
\delta \phi \sim r_c^{-1-\gamma }r_g^{-1/2}r^{3/2+\gamma }.
\end{equation}
\label{precgamma}
Observations of accuracy $\sigma_ \phi$ can therefore test gravity
theories with
\begin{equation}
r_c < r \left( {r\over \sigma _\phi ^2r_g}\right) ^{1\over
2(1+\gamma )}.
\end{equation}

One can speculate that in the absence of additional scales 
the gravity theories that produce self-acceleration, without vacuum energy should have
 $r_c\sim$few Gpc. Then the lunar precession accuracy of $\sigma _\phi \sim 10^{-12}$ will tests theÊ $\gamma =0$ theory\cite{dgp}. The dependence
of the right-hand side of (\ref{precgamma}) on $\gamma$ is very strong, and
cosmologically interesting theories with $\gamma <0$ are ruled out by
current observations, while theories withÊ $\gamma >0$ are not testable
by the solar system observations.

\section{vDVZ Discontinuity and its Absence}

 Some time ago van Dam and Veltman, and Zakharov \cite{vdvz} suggested that the solar system
observations rule out the possibility of non-zero  graviton mass, no matter how small.
Their conclusion was based on a linearized theory of massive graviton with the following action 
\begin{equation}
S \, = \,  S_{El}  \, + \,  \int \, d^4x \, \left ( {M_{Pl}^2 \, m_g^2 \over 2} \, (\, h_{\mu\nu}^2 \, - \, (h_{\mu}^{\mu})^2) \, 
+ \, {1 \over 2} h_{\mu\nu} T^{\mu\nu} \right ) 
\label{linear}
\end{equation}
The first term on the r.h.s. is the standard Einstein action expanded to a quadratic order in
the metric fluctuations about the flat space $g_{\mu\nu} \, = \, \eta_{\mu\nu} \, + \, h_{\mu\nu}$. 
The mass term in (\ref{linear}), has the Pauli-Fierz form \cite{pf}, which is the only possible ghost-free 
combination  quadratic in $h_{\mu\nu}$. 

 Discontinuity is due to the fact that massive graviton contains five degrees of freedom 
(five polarizations)  as opposed to two polarizations  in the massless case. Three out of the five
polarizations couple to the conserved energy momentum source, leading to an additional scalar
attraction at distances $r \, \ll \, {1 \over m_g}$, as compared to the massless case.  Since the additional degree of freedom couples differently to relativistic and non-relativistic sources, the effect
is not merely reducible to the rescaling of the Newton's constant $G_N$, and is observable
at the level of  one-graviton exchange. Thus, at this level the theory is discontinuous in the limit
$m_g \, \rightarrow \, 0$.

The amplitude
of the lowest tree-level exchange by a single massless graviton  between two 
sources with energy-momentum tensors $T_{\mu\nu}$
and  $T^{\prime}_{\alpha\beta}$  is (the tilde sign denotes the quantities
which are Fourier transformed to momentum space):
\beq
{\cal A}_{massless}\,
= \,- \frac{8\pi\,\gn}{q^2}\left 
( {\tilde T}_{\mu\nu}\,-\,{1\over 2}\,
\eta_{\mu\nu} 
\,{\tilde T}^\beta_\beta   \right )\, {\tilde T}^{\prime\mu\nu}\,.
\label{4DT}
\eeq
In the massive case this amplitude takes the form:
\beq
{\cal A}_{massive}\,
=\, -\frac{8\pi\,\gn}{q^2+m_g^2}\left ( {\tilde
T}_{\mu\nu}\,-\,{1\over 3}\,
\eta_{\mu\nu} 
\, {\tilde T}^\beta_\beta   \right )\, {\tilde T}^{\prime\mu\nu}\,.
\label{4DTm}
\eeq
The additional scalar attraction for massive case, is reflected in the
difference in the tensor structure.  This difference can not be eliminated by simple redefinition
of parameters and is finite, for arbitrarily small graviton mass.

 This result goes under the name of vDVZ discontinuity, and if true, would be an extremely powerful result, as it would rule out not only the possibility of massive gravity, but much wider classes of theories that modify gravity in far-infrared. 

 However, the story is not so straightforward as we shall now discuss.  Shortly after vDVZ observation, Vainshtein \cite{Arkady} suggested that vDVZ discontinuity was an artifact of the linearized approximation, and would be absent in fully non-linear theory. That is, vDVZ result was obtained at
one-graviton   exchange level, that is in the first order in $G_N$ expansion.
This corresponds to solving the linearized equation
\begin{equation}
{\delta \, S_{El} \over \delta h_{\mu\nu}} \, + \, m_g^2 \, (h^{\mu\nu} \, -\, \eta_{\mu\nu}h_{\alpha}^{\alpha})\,
= \, {T^{\mu\nu} \over M_{Pl}^2}
\label{eqlinear}
\end{equation}
However, Vainshtein noted that perturbative expansion in $G_N$ breaks down in 
the zero $m_g$ limit. 
However, he also showed that if one takes into account non-linearities, then 
for small $m_g$ the perturbative expansion in powers of $m_g$ can be organized, in which case
discontinuity can disappear. Since no fully non-linear generally-covariant  theory
for massive graviton was known, Vainshtein used a theory obtained by non-linear completion of
only the first, Einstein,  term in r.h.s. of (\ref{eqlinear}):
\begin{equation}
\label{arkequation}
G_{\mu\nu}  \, + \, m_g^2 (h_{\mu\nu} -\eta_{\mu\nu}h_{\alpha}^{\alpha})\, = \, 8\pi G_N \,
T_{\mu\nu} 
\end{equation}
where $G_{\mu\nu}$ is the Einstein tensor.
In which case solving for  $g_{00} \, = \, {\rm e}^{\nu(r)}$ and $g_{rr}\, = \, {\rm e}^{\lambda(r)}$ components of the spherically-symmetric metric outside massive body,
he found:
\begin{eqnarray}
&&\nu(r) \,=  \,-{r_g\over r}+{\cal O}\left(m_g^2\sqrt{r_g
r^3}\right)\,,\qquad
\lambda(r) \,= \,{r_g\over r}+{\cal O}\left(m_g^2\sqrt{r_g
r^3}\right)\,,\nonumber\\[1mm]
\label{mg}
\end{eqnarray}
Note that in the same parametrization  the standard Schwarzschild solution 
of the massless theory takes the following form:
\vspace{0.5cm}
\begin{eqnarray}
&&\nu^{\rm Schw}(r) \,=\, -
\lambda^{ \rm Schw}(r)\,=\, \ln\left(1- {r_g\over
r}\right)\,=\, - {r_g\over r}- \frac 1 2 \left( {r_g\over
r}\right)^2+\ldots\;,\nonumber\\[1mm]
\label{Schw}
\end{eqnarray}
Here $r_g\equiv 2\gn M$ is the gravitational radius of the 
source of mass $M$. 

This expression is fully continuous in $m_g$ and reproduces Einsteinian results in the zero-mass limit.
 Thus  vDVZ has indeed disappeared at the non-linear level.
Notice the two important facts. First the sub-leading corrections to the metric are non-analytic in
$r_g$. Secondly, the  deviations from the standard Einstein gravity become important
at distance
 \beq
 r_*={(m_g\, r_g)^{1/5}\over m_g }\, ,
\label{int}
\eeq
which is parametricaly shorter than the Compton wave-length of graviton.  This is not surprising,
if we recall that vDVZ was cured by non-linearities that are most important in the neighborhood of
the heavy sources (large $r_g$).  As a result the heavier is the source, the larger is the critical distance
at which linear approximation takes over.  

 The above results however, may not sound
completely satisfactory, since the mass term in (\ref{arkequation}) was still kept at the linearized level, and as a result
theory was not fully generally-covariant.  As we shall see,  the above results nevertheless
persists  in theories, which are fully non-linear and generally-covariant.  There too the
discontinuity is absent and  corrections can penetrate at distances much shorter than the scale at which linearized gravity gets modified. The latter fact gives possibility of experimentally testing 
such theories through precision measurements at the relatively short distances.

An example of generally-covariant theory that modifies gravity in far infrared,
and which exhibits vDVZ at the linearized level is given by the following action\cite{dgp,dg}
\beq
S~=~{\Mpl^2 \over 4r_c} ~ \int d^4x ~dy~ \sqrt{|g^{(5)}|}~ \Rbu 
 +~
{\Mpl^2 \over 2}~ \int d^4x ~ \sqrt{|g|}~\Rbr(x)~. 
\label{action}
\eeq
Where $g_{AB}^{(5)}$ is 5D metric tensor, $A,B$
are five-dimensional indexes,  and $\Rbu$ is the five-dimensional Ricci scalar, 
$g_{\mu\nu}$ denotes the induced metric on the brane
which we take as 
\beq
g_{\mu\nu} (x)~\equiv~g^{(5)}_{\mu\nu}(x, y=0)~,\qquad \mu,\nu=0,1,2,3~, 
\label{gind1}
\eeq
neglecting the brane fluctuations.

We assume that our observable 4D world 
is confined to a  brane which is located  at the point 
$y=0$ in extra fifth dimension.  That is  the energy-momentum tensor 
of 4D matter has the form 
$T_{\mu\nu}(x)\,\delta(y)$. 

Although, the underlying  theory is high-dimensional, from the point of view of a 4D observer
localized on the brane, Newtonian gravitational potential is just the usual $1/r$ gravity, which gets modified to $1/r^2$ only at very large distances $r \gg r_c$. 

 The gravitational potential between the two bodies located on the brane can be read-off
from the form of the Greens function (it is convenient to
 work in momentum space in the four
world-volume directions and in position space with respect to
the transverse coordinate  $y$).

Neglecting
the tensorial structure of the propagator
the scalar part of the Green function has the following form\cite{dgp}:
\beq
{\tilde G}(q,y=0)\,=\,{1\over \Mpl^2}~
{1\over q^2 \,+ \,  {\sqrt{q^2} \over r_c}}\, ,
\label{prop}
\eeq
For the static gravitational potential $\Psi(r)$, one gets 
at short distances, i.e.,  when $r\ll r_c $
\beq
\Psi(r)\,=\,-{1\over 8\pi^2 \Mpl ^2}\,{1 \over r}\,\left \{
{\pi\over 2} +\left [-1+\gamma -{\rm ln}\left ( {r_c\over r } \right ) 
\right ]\left ( {r\over r_c } \right )\,+\,{\cal O}(r^2)  
\right \}\,.
\label{short}
\eeq
Here $\gamma\simeq 0.577$ is the Euler constant.
The leading term in this expression has the familiar $1/r$ scaling of the
four-dimensional Newton  law.
 For $r\gg r_c $ one finds:
\beq
\Psi(r)\,=\,-{1\over 16\pi^2 \M^3}\,{1 \over r^2}\,+
\,{\cal O} \left ( {1\over r^3} \right )\,.
\label{long}
\eeq
The long distance potential scales as $1/r^2$ in accordance 
with the  5D Newton law.

The above  model (\ref {action}) exhibits the vDVZ 
discontinuity in the one-graviton tree-level exchange. 
This can be seen directly from  the $\mu\nu$ components of linearized Einstein equation
for bulk metric fluctuations about the flat
5D metric, which after gauge fixing can be brought into the following form\cite{dgp}
\beq
\left ( {1 \over r_c} (\nabla^2 \, - \,  \partial_y^2)\, + \, \delta(y)
\, \nabla^2
\right )\, h_{\mu\nu} \,  = \, -\, {1 \over M_{Pl}^2} \,\left \{ T_{\mu\nu} -{1\over 3}\eta_{\mu\nu} 
T^\alpha_\alpha   \right \}\delta(y)
+ \delta(y) \,\partial_\mu
\partial_\nu \,h^\alpha_\alpha\,.  
\label{basic}
\eeq
The first term on r.h.s of this equation has a structure which is identical to that of  
a massive 4D graviton.  
The second term,
$\partial_\mu\partial_\nu$,  vanishes vanishes whenever it is contracted with the 
conserved energy-momentum tensor, and thus plays no role at one graviton exchange
level.   As a result, the amplitude of interaction
of two test sources takes the form:
\beq
{\cal A}(q) ~\propto~
{  {\tilde T}^{\mu\nu}{\tilde T}^{\prime}_{\mu\nu}~-~
{1\over 3} ~{\tilde T}^{\mu}_\mu
{\tilde T}^{\prime\nu}_\nu \over q^2~+~\, {q \over r_c} }~,
\label{prop1}
\eeq  
where $q\equiv \sqrt{q^2}$. 
We see that the tensor structure 
is the same as in the case of the massive 4D theory, 
see Eq. (\ref {4DTm}), which signals vDVZ discontinuity. 

The discontinuity however is cured at the nonlinear level. This was demonstrated explicitly both
for cosmological solution\cite{ddgv}, as well as for cosmic strings\cite{lue} and 
Schwarzchild\cite{gruzinov},\cite{masimo}. We shall concentrate on the latter case.
The solution of \cite{gruzinov} has the form:
\begin{eqnarray}
&&\nu(r) \,=  \,-{r_g\over r}+{\cal O}\left({1 \over r_cr}\sqrt{r_g
r^3}\right)\,,\qquad
\lambda(r) \,= \,{r_g\over r}+{\cal O}\left({1 \over r_cr}\sqrt{r_g
r^3}\right)\,,\nonumber\\[1mm]
\label{andrei2}
\end{eqnarray}
This expression exhibits some interesting behavior.  
First, we see that the two features observed by Vainshtein persist here, but there are some peculiarities. The correction is non-analytic
in $r_g$, and moreover it becomes important  at distances much closer than $r_c$.  This is in 
accordance with Vainshtein case. However, there are important differences which are
crucial for observations. Notice that, one would naively expect that if Vainshtein's analysis is
correct, then  the above expression should be obtained from 
Vaishtein case by substitution $m_g \rightarrow 1/r_c$. The reason is that, although we are not
dealing with massive gravity in the strict sense, nevertheless, $r_c$ is the scale at which
$1/r$-law of linearized approximation breaks down.  It is true that in our case it is replaced by
$1/r^2 $ rather than by exponentially suppressed Youkawa potential, but this difference naively seems  inessential.
 So naively one would conclude that $r_c$ should play the role of $m_g$ in controlling the strength of correction. 
However, this is not the case. Instead, the expression of ref\cite{gruzinov} is obtained from Vainshtein by substitution:
\begin{equation}
m_g^2 \, \rightarrow \, {1 \over rr_c}
\label{ substitution}
\end{equation}

We shall now try to understand why this is the case, but let us first explain non-analyticity of the
correction in $r_g$.  The above theory contains two parameters $M_{Pl}$ and $r_c$.
vDVZ happens at the linear level due to the term that is proportional to $1/r_c$, and is cured
by the non-linear corrections proportional to $M_{Pl}^2$. These non-linear corrections make sure that metric gets standard Einstein form near the sources. The deviation from the standard metric is due to the fact that linearized approximation becomes good again and  $1/r_c$-terms take over.
Thus, corrections to the metric are due to the fact that one-particle exchange dominates at large scales.
Then the leading corrections should be proportional to
$1/r_c$.  Close to the source the metric can be expanded in powers of $1/r_c$, e.g.
\begin{equation}
g_{00} \,  ( \, r, \, r_c, \, r_g \, ) \, =  \, 1 \, + \, {r_g \over r} \, +  \, {1\over r_c}  \, r_g^{\alpha} r^{\, 1 \, - \, \alpha} \, + \, {\rm higher~order~terms}
\label{correction}
\end{equation}
where $\alpha$ is some positive power.  Now, we know that for the fixed $r$, for weak sources, the
linearized approximation is valid, and thus the effect of extra graviton polarization must be reintroduced. Thus, for fixed $r$, the second term on r.h.s. of (\ref{correction}) must dominate
 in the limit $r_g \rightarrow 0$. This can only happen if $\alpha < 1$, which explains non-analyticity of the correction.  To summarize shortly, given the fact that difference in the tensor structure of the metric is introduced by dominance of one-particle exchanges,
which are proportional to ${1 \over r_c}$, and dominate for small $r_g$, 
the corrections to Schwarzchild cannot be analytic in $r_g$.

To understand which quantity plays the role of $m_g$, the following observation is useful.
If we are interested in making 4D metric on the brane continuously approach 
Schwarzchild in the limit $1/r_c \, \rightarrow 0$, it is enough to have non-linear action on
the brane only, and keep high-dimensional part of the action linear.
In other words for curing
vDVZ the non-linear interactions in the bulk are unimportant. 
That is, for obtaining Schwarzchild solution in the leading order on the  brane the following equation is enough
\beq
 M_{Pl}^2\left ( \, {1\over r_c} \, {\cal G}_{\mu\nu}^L \, + \,  \delta(y) G_{\mu\nu} \, \right ) \,
= \, \delta(y) T_{\mu\nu}
\label{partlinear5}
\eeq  
Where, ${\cal G}_{\mu\nu}^L$ is the five-dimensional Einstein tensor, linearized on a flat background.
This system describes the five dimensional graviton $h_{\mu\nu}(x,y)$ 
 that freely propagates in the bulk, and has non-linear self-couplings only on the brane.  These brane-localized self couplings are most important for
determining the metric on the brane near the source localized on the brane.  Diagrammatically,
this fact can be understood as follows. The  nonlinear corrections that cure vDVZ discontinuity correspond to
all the tree-level diagrams with virtual graviton lines  that end on the sources localized on the brane. The self-interaction vertexes of virtual gravitons can be located both on the brane or
in the bulk. The bulk non-linearities come from the $1/r_c$-suppressed bulk curvature term and are subleading.  So the
diagrams for which graviton self-interaction vertices are not located on the brane are sub-leading in $1/r_c$-expansion. Theory with linearized bulk action (\ref{partlinear5}) gives rise only to the diagrams in which all graviton vertices are located
on the brane. Since these diagrams are dominant, the eq (\ref{partlinear5}) is suffices  to cure
vDVZ discontinuity.
Now let us compare the five-dimensional theory defined in by eq. (\ref{partlinear5}) to 
an effective four-dimensional theory of four-dimensional graviton
$h_{\mu\nu}(x)$ defined by the following equation
\beq
 M_{Pl}^2 \left ( \, G_{\mu\nu} \, +  \,  {1\over r_c} \sqrt{\nabla^2} (h_{\mu\nu} \, - \, \eta_{\mu\nu} h_{\alpha}^{\alpha})\,\right) \,  = \, T_{\mu\nu}
\label{partlinear4}
\eeq
These can be treated as two independent theories, but some sub-class of amplitudes in the
two cases are equal.  In fact the interaction amplitudes between brane-localized sources
with no graviton emission in  5D theory are equal to the ones of 4D one. 
Notice that the two theories are designed in such a way, that the Greens function of
the four-dimensional theory, is equal to the greens function of the five-dimensional one
evaluated at the point $y=0$:
\begin{equation}
\label{4and5 }
\langle \, h_{\mu\nu}(x)\, h_{\alpha\beta}(x') \, \rangle \, = \, \langle 
\, h_{\mu\nu}(x, \, y\, = \, 0)\, h_{\alpha\beta}(x', \, y \, = \, 0)\, \rangle
\end{equation}
Also all the nonlinear interactions in two theories are the same.
Given these facts, it is obvious that any interaction amplitude between the  brane-localized
sources $T_{\mu\nu}(x_1), \, T_{\mu\nu}(x_2), ...\,T_{\mu\nu}(x_n), \,$ (in which no gravitons are emitted in the final state) in five dimensional theory
(\ref{partlinear5}) will be equal to a similar  amplitude in four-dimensional theory
(\ref{partlinear4}). For instance consider the lowest  three-level interaction among the three sources with a single intermediate three-graviton vertex $V(x)\delta(y)$:
\begin{eqnarray}
\label{3amplitude}
&&{\cal A} \sim \int \, d^4x_1dy_1d^4x_2dy_2d^4x_3dy_3d^4xdy\, 
T^{\mu\nu}(x_1)\delta(y_1) \, T^{\alpha\beta}(x_2) \delta(y_2) \,T^{\gamma\rho}(x_3) \, \delta(y_3)
\, \nonumber\\
&&V^{\mu'\nu',\alpha'\beta',\gamma'\rho'}(x) \, \delta(y) \,
\langle \, h_{\mu\nu}(x_1 \, y_1)\, h_{\mu'\nu'}(x, \, y )\, \rangle \,
\langle \, h_{\alpha\beta}(x_2, \, y_2\, )\, h_{\alpha'\beta'}(x, \, y )\, \rangle \,
\nonumber\\
&&\langle \, h_{\gamma\rho}(x_3, \, y_3)\, h_{\gamma'\rho'}(x, \, y )\, \rangle \,
 = \, \nonumber\\
&&\int \, d^4x_1d^4x_2d^4x_3d^4x\, 
T^{\mu\nu}(x_1) \, T^{\alpha\beta}(x_2) \,T^{\gamma\rho}(x_3) \,
 V^{\mu'\nu',\alpha'\beta',\gamma'\rho'}(x) \, 
\langle \, h_{\mu\nu}(x_1)\, h_{\mu'\nu'}(x)\, \rangle \, 
\nonumber\\
&&\langle \, h_{\alpha\beta}(x_2)\, h_{\alpha'\beta'}(x)\, \rangle \,
\langle \, h_{\gamma\rho}(x_3)\, h_{\gamma'\rho'}(x)\, \rangle
\end{eqnarray}
The last expression represents a similar amplitude evaluated in theory of eq. (\ref{partlinear4}). 
Generalization of the above relation to more complicated diagrams with arbitrary number of
internal vertexes and external sources is trivial.  As a result  any such amplitude in high-dimentional theory eq. (\ref{partlinear5}) has an equal counterpart in  four-dimensional one.  
Thus, if for example,  we are interested in the metric created by the brane-localized source on the brane in theory defined by eq. (\ref{partlinear5}),  we can instead solve eq. (\ref{partlinear4}) and get the correct answer. 
But for solving perturbatively for  Schwarzschild metric, the equation (\ref{partlinear4}) in
the leading order is equivalent to that of  Vainshtein (\ref{arkequation}) in which
$m_g^2$ is substituted by $1/rr_c$.

Now we can generalize this result and give a simple qualitative prescription that gives a possibility to estimate the sub-leading correction to Schwarzschild in
other theories of interest,  in which Einstein gravity is modified in far infrared by adding some
operators that at the linearized level behave as
\begin{equation}
G_{\mu\nu}  + {1 \over r_c^2}  f(r_c^2\nabla^2) (h_{\mu\nu} \, - \, \eta_{\mu\nu}\, h_{\alpha}^{\alpha})\,  = \, 0
\label{ modified}
\end{equation}
Then subleading correction to the metric can be estimated from (\ref{mg})  by substitution
\begin{equation}
m_g^2 \rightarrow {1 \over r_c^2} f \left ({r_c^2 \over r^2} \right )
\label{substitution}
\end{equation}
where action of the operator should be understood in terms of the eigenvalues of $\nabla$.
For instance:
\begin{equation}
f(r_c^2\nabla^2)  \, {1 \over r} \, = \, \int \, d^3p \, f(r_c^2p^2) \, {{\rm e}^{-irp}  \over p^2}
\label{actionf }
\end{equation}
It should be noted that we have neglected corrections of the same order in $1/r_c$ coming from the nonlinear completion of the mass term. Thus our results is just an order of magnitude estimate. In principle there could be theories in which the additional terms exactly cancel the corrections to Schwarzschild we computed although this is not the case in the theories of \cite {dgp,dg}.

\section{Applicability}

 We shall now briefly formulate the applicability range of our analysis. Consider a theory which,
in some gauge, at the linearized level reduces to eq.  (\ref{modf}).
The criteria for the presence of discontinuity ($\beta \neq 1$) is the
existence of the following spectral representation:
\begin{equation}
\label{spectral}
{1 \over q^2 \, + \, f(r_c^2q^2)} \, = \, \int _0^{\infty} \, {ds \, \rho(s) \over s + q^2}
\end{equation}
with a semi-positive definite spectral function $\rho(s)$, which is requirement for the absence of
unphysical negative norm states. In such a case the equation (\ref{modf}) is derivable from the
action
\begin{equation}
S \, = \int_0^{\infty} \, ds \left [ \, S_{El}^{(s)}  \, + \,  \int \, d^4x \, \left ( {s \over 2} \,(\, (h^{(s)}_{\mu\nu})^2 \, - \,  (h^{(s)\mu}_{\mu})^2) \, 
+ \, {\sqrt{\rho(s)} \over 2 M_{Pl}} h^{(s)}_{\mu\nu} T^{\mu\nu} \right )\, \right ]
\label{slinear}
\end{equation}
where $h_{\mu\nu} =\int ds \sqrt{\rho(s)} h^{(s)}_{\mu\nu}$.
This action describes the linearized theory of continuum of free massive gravitons with masses
$m^2\,  = \, s$. They all couple to the same conserved source $T_{\mu}^{\nu}$ but with mass-dependent
coupling $\sqrt{\rho(s)} \over 2 M_{Pl}$. 
Therefore the propagator of this theory is simply an integral over the continuum of the
massive  propagators taken with the weights $\rho(s)$
\begin{equation}
\label{samplitude}
{\cal A}(q) \, = \, - \,{8\pi\,\gn} \int_0^{\infty} \, ds \rho(s) { {\tilde
T}_{\mu\nu}\,{\tilde T}^{\prime\mu\nu} \, - \, {c(s) \over 3} \,
\, {\tilde T}^\mu_\mu \, {\tilde T}_{\nu}^{\prime\nu}\, \over s+q^2}
\end{equation}
where in our parametrization $c(0) \, = \,3/2$ and $c(s \neq 0)\, = \,1$.

A useful illustrative example, in which vDVZ can be seen from the above reasoning is
the model of \cite{dgp}. There the representation (\ref{slinear}) has a simple physical 
meaning \cite{dgkn}. It is just an expansion into the {\it continuum} of massive Kaluza-Klein states.  In that case
the spectral function is given by
$\rho (s)\, = \, {1 \over \sqrt{s} (4 \, + \, r_c^2s)}$. 
In this language the presence of vDVZ discontinuity in \cite{dgp} is trivially
understood. Since each Kaluza-Klein mode is a massive spin-2 particle, each of them 
exhibits vDVZ discontinuity by default. Thus exchange by tower of Kaluza-Klein states exhibits the same discontinuity as the individual exchanges do.  

\section{Conclusion}

There are strong motivations coming from cosmology for modifying the standard Einstein gravity at large distances. A wide class of such gravity theories can be tested by astronomical observations of the solar system.

\vspace{0.5cm}   

{\bf Acknowledgments}
\vspace{0.1cm} \\

We thank  Gregory Gabadadze and  Glennys Farrar for valuable discussions
and  Erich Adelberger for the discussions on Lunar Ranging experiments, that inspired this work.
 The work of G.D. is supported in part 
by David and Lucile  
Packard Foundation Fellowship for  Science and Engineering,
by Alfred P. Sloan foundation fellowship and by NSF grant 
PHY-0070787.  Work of A.G. is supported by part by David and Lucile  
Packard Foundation Fellowship for  Science and Engineering.
Work of M. Z. is supported by part by David and Lucile  
Packard Foundation Fellowship for  Science and Engineering.

\end{document}